\documentclass[acmsmall,screen]{acmart}

\usepackage{multirow}
\usepackage{listings}
\usepackage{url}

\AtBeginDocument{%
  \providecommand\BibTeX{{%
    \normalfont B\kern-0.5em{\scshape i\kern-0.25em b}\kern-0.8em\TeX}}}

\begin{document}

\title[Short title]{Social Diversity for ATL Repair}


\author{Zahra VaraminyBahnemiry}
\email{varaminz@iro.umontreal.ca}
\affiliation{%
  \institution{Université de Montréal, DIRO}
  \country{Canada}
}

\author{Jessie Galasso}
\email{jessie.galasso-carbonnel@umontreal.ca}
\affiliation{%
  \institution{Université de Montréal, DIRO}
  \country{Canada}
}

\author{Houari Sahraoui}
\email{sahraouh@iro.umontreal.ca}
\affiliation{%
  \institution{Université de Montréal, DIRO}
  \country{Canada}
}

%\renewcommand{\shortauthors}{Short authors}
%\acmConference[ESEC/FSE 2022]{The 30th ACM Joint European Software Engineering Conference and Symposium on the Foundations of Software Engineering}{14 - 18 November, 2022}{Singapore}%%

\begin{abstract}
Model transformations play an essential role in the Model-Driven Engineering paradigm. Writing a correct transformation program requires to be proficient with the source and target modeling languages, to have a clear understanding of the mapping between the elements of the two, as well as to master the transformation language to properly describe the transformation. Transformation programs are thus complex and error-prone, and finding and fixing errors in such programs typically involve a tedious and time-consuming effort by developers. In this paper, we propose a novel search-based approach to automatically repair transformation programs containing many semantic errors. To prevent the \textit{fitness plateaus} and the \textit{single fitness peak} limitations, we leverage the notion of social diversity to promote repair patches tackling errors that are less covered by the other patches of the population. We evaluate our approach on 71 semantically incorrect transformation programs written in ATL, and containing up to five semantic errors simultaneously. The evaluation shows that integrating social diversity when searching for repair patches allows to improve the quality of those patches and to speed up the convergence even when up to five semantic errors are involved.

\end{abstract}

%\begin{CCSXML}
%<ccs2012>
%  <concept_id>10010520.10010553.10010562</concept_id>
%  <concept_desc>Computer systems organization~Embedded systems</concept_desc>
%  <concept_significance>500</concept_significance>
%  <concept_id>10010520.10010575.10010755</concept_id>
%  <concept_desc>Computer systems organization~Redundancy</concept_desc>
%  <concept_significance>300</concept_significance>
%  <concept_id>10010520.10010553.10010554</concept_id>
%  <concept_desc>Computer systems organization~Robotics</concept_desc>
%  <concept_significance>100</concept_significance>
%  <concept_id>10003033.10003083.10003095</concept_id>
%  <concept_desc>Networks~Network reliability</concept_desc>
%  <concept_significance>100</concept_significance>
%</ccs2012>
%\end{CCSXML}

%\ccsdesc[500]{Computer systems organization~Embedded systems}


\maketitle

\section{Introduction}

Model-driven engineering (MDE) has been encouraged for many years as an efficient approach to reduce software development complexity through increasing the abstraction level ~\cite{Mohagheghi:2013}. In this context, MDE sees models as first class artifacts. 
It combines domain specific modeling languages to capture specific aspects of the solution, and transformation engines in order to produce from these models low level artifacts such as source code, documentation and test suites~\cite{schmidt2006model}.
Model transformations effect an essential role in the MDE approach.
Model transformation programs depict how to transform elements of a model conforming to a source meta-model into elements of a model conforming to a target meta-model.

They can be written with general purpose programming languages or in dedicated transformation languages such as DSLTrans~\cite{barroca2010dsltrans} or the ATLAS Transformation Language (ATL)~\cite{jouault2008-ATL}.
Writing a correct model transformation program requires to be proficient with the source and target meta-models, to have a clear understanding of the mapping between the elements of the two and to know how to exploit the transformation mechanisms of the language to properly describe this transformation.
Transformation programs are thus complex and error-prone, and finding and fixing errors in such programs typically involve a tedious and time-consuming effort by developers.   

Several types of errors can affect a transformation program.
Syntactic errors usually prevent the transformation from compiling and producing an output model.
To alleviate the developers' effort when fixing syntactic errors, works such a the one of Cuadrado~\cite{cuadrado2018-quickfix} propose predefined corrective patches to be applied on errors detected with syntactic analysis tools (e.g., AnATLyzer~\cite{cuadrado2018-anatlyser} for the ATL language).
When the transformation program compiles but the implemented behavior is not the one that was intended by the developers, we say that it contains semantic errors.  
As a consequence, semantically incorrect transformations can produce output models, but the latters are different from the ones that were expected.
Because semantic errors pertain to the transformation's behavior and each faulty transformation needs tailored patches, predefined patches are not suited for semantic errors.

Population-based evolutionary algorithms (EAs) have been widely used to correct errors in programs~\cite{monperrus2018-progrepairbiblio}, including both syntactic~\cite{DBLP:journals/corr/abs-2012-07953} and semantic~\cite{DBLP:conf/models/VaraminyBahnemiry21} errors in transformation programs. 
Formulating program repair as an optimization problem enables such search-based approaches to find patches that will fix a given faulty program in the space of all possible patches.
EAs maintain a population of candidate patches which undergo a process of evolution across several generations until an optimal patch is found.
At each generation, the evolution process creates new solutions based on the population of the previous generation, and the best candidates are retained for the next, hopefully better, generation.
Finding suitable patches with this approach is a fully automated process, at the end of which, the best fitting patches can be presented to the expert to make a final decision about the repair to be applied.
To fix errors related to a program's behavior,  automated approaches usually rely on a specification of the expected behavior (e.g., test cases or examples) to assess the fitness of a patch, and thus efficiently guide the search strategy.

In a previous work~\cite{DBLP:conf/models/VaraminyBahnemiry21}, we used EAs with test cases to correct semantic errors in ATL transformation programs.
We found that this approach usually finds patches to correct transformations having fewer errors, but either have trouble or take too much time to converge toward good patches in the presence of more errors.
Preliminary analysis showed that using test cases to assess the fitness of the corrective patches make the search space difficult to explore efficiently due to \textit{fitness plateaus}~\cite{de2018novel}, an issue of EAs which impede the ability of the approach to converge toward optimal patches.
In addition, EAs are known to give more power to good solutions, which can cause converging issues due to loss of diversity, a problem known as \textit{single fitness peak}.
Using behavior specifications such as test cases to guide the search in EAs can exacerbate these limitations~\cite{batot2018injecting,de2018novel}.

In this paper, we propose a new approach based on EAs to automatically find patches correcting many semantic errors which limits the problems related to convergence. 
To improve the efficiency and effectiveness of EAs using test cases, we propose to leverage the notion of social diversity to promote patches which tackle errors that are less covered by the other patches of the population. 
Our hypothesis is that including this measure in the process will maintain a certain level of diversity and reduce the negative impact on convergence of single fitness peak and fitness plateaus.
To include this notion in EAs, we formulate the transformation repair as a multi-objective optimization problem, where solutions must optimize several objectives. 
Our approach is implemented using the NSGA-II algorithm, a fast multi-objective EA~\cite{deb2000-nsga}.

We perform an evaluation on 71 semantically incorrect transformation programs written in ATL to assess the impact of social diversity on the convergence of EA-based repair.
We reuse the faulty transformations from our previous work and also consider new transformations taken from the ATL zoo\footnote{\url{https://www.eclipse.org/atl/atlTransformations/}} containing up to 5 semantic errors to thoroughly evaluate the impact of our approach on transformations having many errors.
The evaluation shows that social diversity is able to improve both the efficiency and the efficacy of EAs to fix faulty transformation programs, even when they contain up to 5 semantic errors.

In Sect.~\ref{sec:background}, we briefly describe ATL transformation programs and give examples of defects in those transformations. 
We also provide examples of patches repairing such defects.
Section~\ref{sec:problemstatement} presents evolutionary algorithms and illustrate how to use them to repair semantic errors in transformation programs.  We also discuss two limitations faced by this approach, related to the use of test cases.
Section~\ref{sec:ourapproach} describes our multi-objective approach using social diversity to automatically generate patches which attempt to overcome these limitations.
The impact of breathing social diversity in EAs to improve convergence and repair many errors is evaluated in Section~\ref{sec:experiments}.
Section~\ref{sec:relatedwork} presents related work and Sect.~\ref{sec:conclusion} concludes the paper.

\section{Background}
\label{sec:background}

In this section, we  first provide background about model-to-model transformation programs.
We focus on programs written in the ATLAS Transformation Language (ATL)~\cite{jouault2008-ATL}, but the approach presented in this paper is generic and can be adapted to other  transformation languages.
We then discuss the types of errors that can be found in such programs, including semantic errors, which are the target of this work.
Finally, we show what is a patch to repair faulty transformation programs.

\subsection{ATL Transformation Programs}

In this section, we provide some background about model transformation programs written in ATL and illustrate the presented notions on an example inspired from the \textit{Class2Relational}\footnote{\url{https://www.eclipse.org/atl/atlTransformations/\#Class2Relational}} transformation of the ATL zoo.

\begin{figure}[ht]
    \centering
    \includegraphics[width=.85\linewidth]{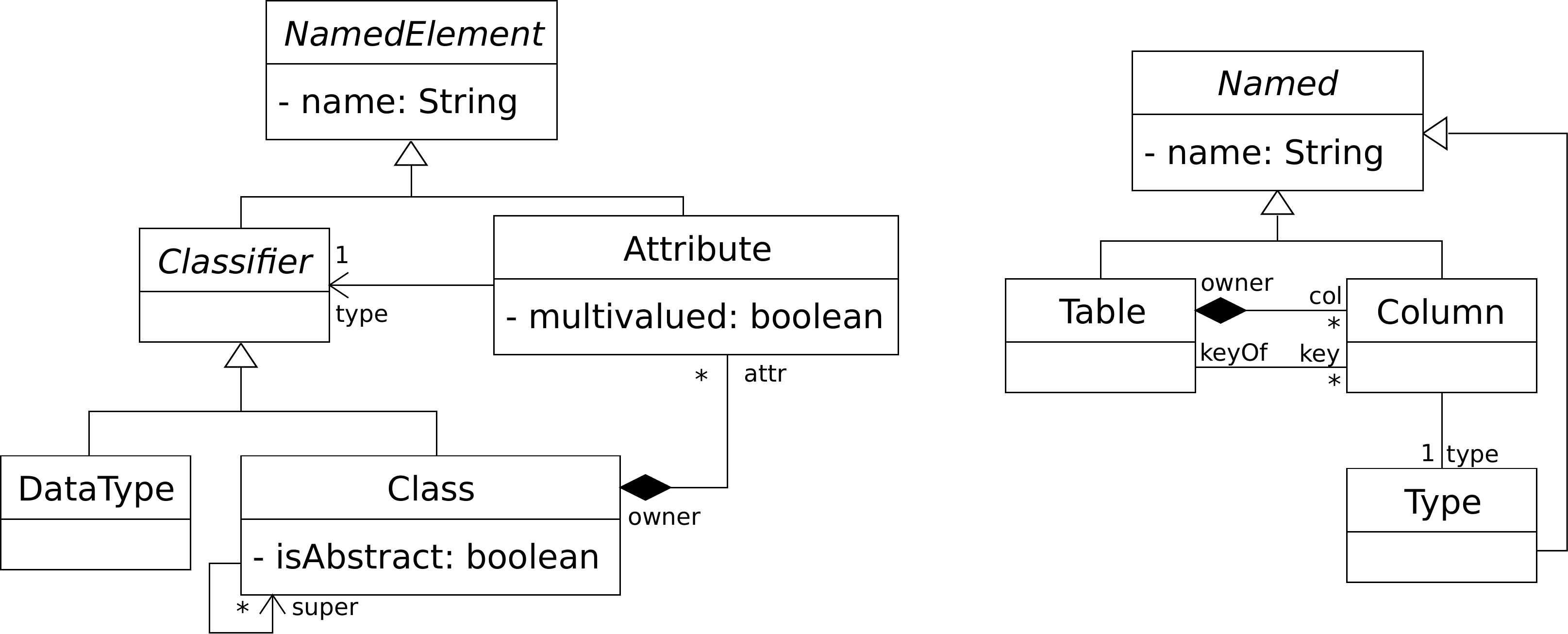}    \caption{UML Class Diagram meta-model (left-hand side) and Relational Schema meta-model (right-hand side) }
    \label{fig:metamodels}
\end{figure}

Model transformation programs automate the process of transforming a source model into a target model.
A transformation program relies on meta-models describing both the source and the target models.
The source and target models can conform to the same meta-model (endogeneous transformation), but in the case of exogeneous transformations, transformation programs usually rely on two different meta-models.
Thus, a given transformation program is defined for a pair of meta-models, and can only transform source models conforming to the input meta-model into a target model conforming to the output meta-model.
For instance, the \textit{Class2Relational} transformation program transforms an UML class diagram into its equivalent relational schema. Figure~\ref{fig:metamodels} shows simplified versions of the UML Class Diagram meta-model and the Relational Schema meta-model as presented in the ATL Zoo Class2Relational transformation. 

Listing~\ref{lst:semanticerrors} presents a simplified excerpt of the \textit{Class2Relational} transformation program written in ATL.
The two meta-models are identified on line 1: the source meta-model is Class (\texttt{IN}) and the target meta-model is Relational (\texttt{OUT}).
A transformation typically consists of a set of rules defining how to transform elements of the input meta-model into elements of the output meta-model.
A rule is introduced by the keyword \texttt{rule} (lines 3, 13, and 20) and has a name (for instance, Class2Table in line 3).
Each rule has two parts: a \texttt{from} part outlining elements of the source meta-model, and a \texttt{to} parts defining how to transform elements of the \texttt{from} part into elements of the target meta-model.
The \texttt{from} part defines a pattern in the source model.
Patterns can represent types: for instance on line 4, the rule \textit{Class2Table} applies to each element conforming to the type \textit{Class}.
They can also be refined with constraints, here defined with the OCL language: for instance, lines 14-16 states that the rule \textit{SingleValuedDataTypeAttribute2Column} applies on elements conforming to the type \textit{Attribute}, having an attribute \textit{type} representing a native type, and an attribute \textit{multiValued} being false.

\begin{lstlisting}[
    breaklines=true,
    keepspaces=false,
    breakindent=0pt,
  % basicstyle=\ttfamily\footnotesize\scriptsize,
    basicstyle=\ttfamily\footnotesize,
%    numbers=left,
    caption={Excerpt of an ATL transformation program, from Class  Diagram to Relational Schema },
    label={lst:semanticerrors}]
1 create OUT : Relational from IN : Class;
2
3 rule Class2Table { 
4 from c: Class!Class 
5 to   out: Relational!Table ( 
6         name <- c.name, 
7         col <- Sequence {key} -> excluding(
8            c.attr->collect(e | not e.multiValued)), 
9          key <- Set {key}), 
10      key: Relational!Column ( 
11         name <- c.name + 'Id)}
12           
13 rule SingleValuedDataTypeAttribute2Column { 
14 from a: Class!Attribute ( 
15         a.type.oclIsKindOf(Class!DataType) 
16         and not a.multiValued) 
17 to   out: Relational!Column ( 
18         name <- a.name)} 
19          
20 rule MultiValuedClassAttribute2Column { 
21 from a: Class!Attribute ( 
22         a.type.oclIsKindOf(Class!Class) 
23         and a.multiValued) 
24 to   out: Relational!Table ( 
25         name <- a.owner.name + '_' + a.name, 
26         col <- Sequence {id, foreignKey}),
27      foreignKey1: Relational!Column ( 
28         name <- a.owner.name.firstToLower() + 'Id'),
29      foreignKey2: Relational!Column ( 
30         name <- a.type + 'Id')}
\end{lstlisting}

The \texttt{to} part describes how to create elements of the target model based on the elements of the source model matching the associated \texttt{from} part.
The \texttt{to} part may create one element (e.g., in lines 17-18, a Column is created when an Attribute is matched) or several ones (e.g., in line 5 and line 10, both a Table and a Column are created when a Class is matched).
For each created element, one can define \textit{bindings} to associate values to the attributes of the created element.
Bindings can use values of the source model elements matched in the \texttt{from} part to initialize the target model elements.
For instance, in line 6, the attribute \textit{name} of Table is initialized using the name of the matched \textit{Class} (i.e., \textit{c.name}).
Bindings can define collections (e.g., a Sequence in line 7, a Set in line 9) and may use iterator or operation calls in initialization (e.g., \textit{firstToLower()} in line 28).

\begin{figure}[ht]
    \centering
    \includegraphics[width=\linewidth]{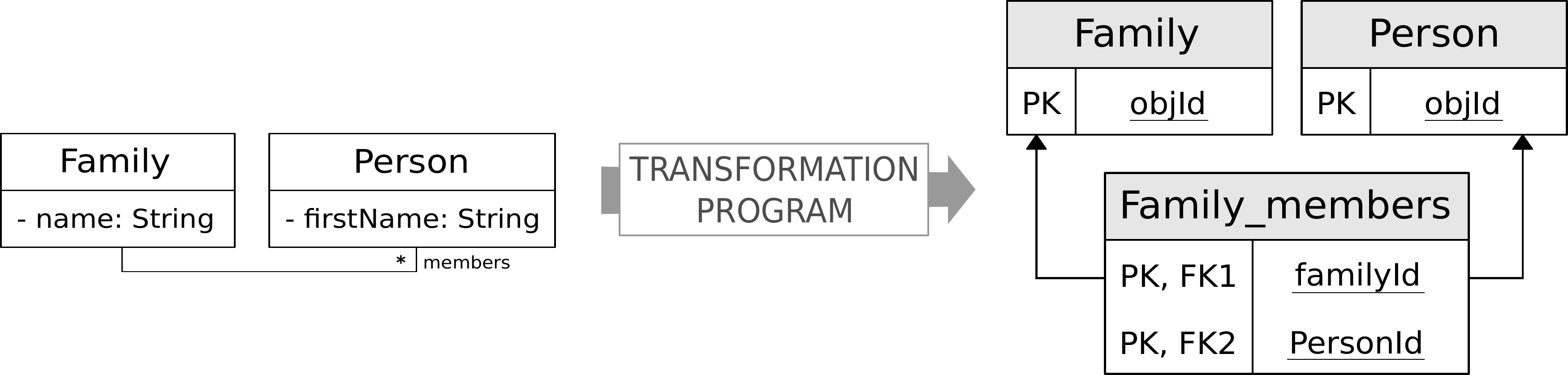}    \caption{Relational Schema (output model, right-hand side) obtained when applying the transformation program of Listing~\ref{lst:semanticerrors} to the class diagram (input model, left-hand side)}
    \label{fig:transformation_example}
\end{figure}

Figure~\ref{fig:transformation_example} shows an example of the target model (right-hand side) conforming to the Relational meta-model obtained when running the transformation of Listing~\ref{lst:semanticerrors} on a source model (left-hand side) conforming to the class diagram meta-model.

\subsection{Defects in Transformation Programs} 

Transformation programs highly depend on the elements of the two meta-models. 
\textit{Syntactic errors} can be due to type misuse such as referring to elements that are not in the meta-models or setting properties with values of the wrong type.
Syntactic errors usually  hinder the proper compilation and execution of the program.
Tools such as AnATLyser~\cite{cuadrado2018-anatlyser}, a static analyzer for the ATL language, can be used to check syntactic errors.

\begin{figure}[ht]
    \centering
    \includegraphics[width=.7\linewidth]{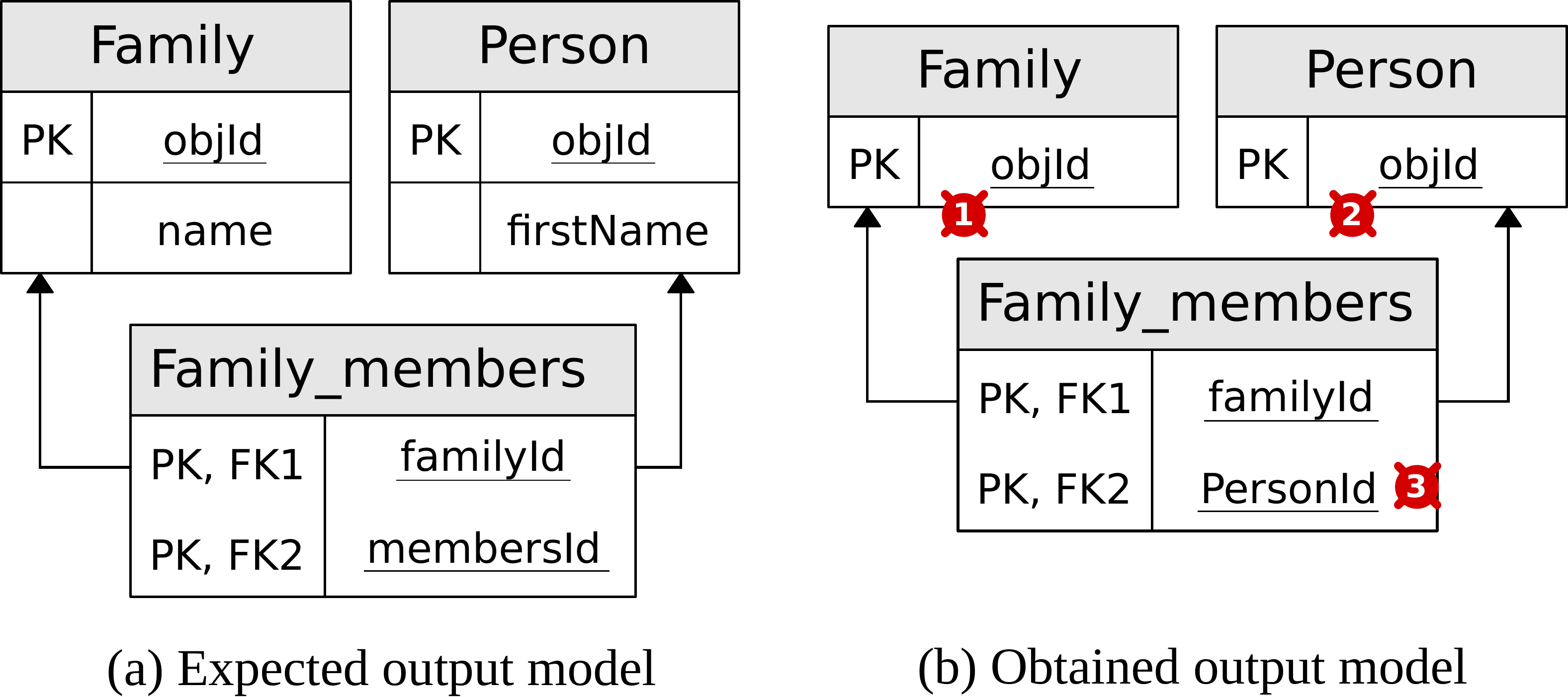}
    \caption{Differences between the expected output model (a) and the output model obtained with the transformation of Listing~\ref{lst:semanticerrors} (b)}
    \label{fig:comp_output_models}
\end{figure}

\textit{Semantic errors} make a program behave in a way that differs from what is expected, i.e., the transformation program is semantically incorrect with respect to a specification of the expected behavior.
These errors do not necessarily hinder the compilation and execution processes, but may cause the program to produce the wrong outputs.
As mentioned earlier, transformation programs are complex and difficult to debug, especially for declarative languages such as 
ATL.
An easy way to outline the intended behavior of a program is to provide a set of inputs-outputs examples of this program, i.e., test cases.
If provided with the source models, the program outputs the expected target models, no behavior deviations (indicating semantic errors) are detected.
However, if the outputted target models are different from those of the provided examples, it shows that the transformation program is semantically incorrect with regards to the provided test cases.
Figure~\ref{fig:comp_output_models} (a) presents the expected target model when providing the transformation program of Listing~\ref{lst:semanticerrors} with the source model of Figure~\ref{fig:transformation_example} (left-hand side).
We can see that it is different from the obtained target model in three different locations, as highlighted in Fig.~\ref{fig:comp_output_models} (b).
The columns \texttt{name} and \texttt{firstName} are missing from the tables \texttt{Family} and \texttt{Person}, respectively.
Also, we can see that the second key of the table \texttt{Family\_members} is named \texttt{PersonId} instead of \texttt{membersId}.
Therefore, according to this test case, Listing~\ref{lst:semanticerrors}  presents a faulty transformation program containing semantic errors.

\subsection{Repair Patches}
\label{sec:example-conformance}

Program repair can be defined by \textit{the transformation of an
unacceptable behavior of a program into a acceptable one according
to a specification}~\cite{monperrus2018-progrepairbiblio}. 

We call a \textit{patch} a sequence of edit operations which modifies a program's source code.
A patch is considered good if it modifies a program to conform to a given specification.
In our case, we consider behavior specifications in the form of pairs of input/output models: if a patch modifies the transformation program so that the obtained output models are equivalent to the expected ones, then this patch is considered optimal to repair the transformation program.

Table~\ref{tab:edit-op} presents a subset of the atomic edit operations for ATL transformations proposed by Cuadrado et al.~\cite{cuadrado2018-quickfix}. 
This subset corresponds to the operations modifying elements in the transformation rules, as presented in~\cite{DBLP:conf/models/VaraminyBahnemiry21}.
\textit{Binding creation} and \textit{binding deletion} respectively add a new and remove an existing binding in a given rule.
\textit{Type of source pattern element} modifies the \texttt{from} part of a rule, while \textit{Type of target pattern element} changes the \texttt{to} part of a rule.
\textit{Type of collection} modifies  collection data types provided by OCL (e.g., Sequences, Set, Bag).
\textit{Type argument of operation} changes the arguments of type-testing operations such as oclIsKindOf() and oclIsTypeof().
\textit{Navigation expression} and \textit{Target of binding} respectively changes a given binding's right-hand side and left-hand side.
Finally, the three operations \textit{Collection operation call}, \textit{Iterator call} and \textit{Predefined operation call} change a call by another.
All these operations take parameters to define on which element it should be applied, as well as the modified values when applicable.
For instance, the edit operation \textit{Navigation expression} considers four parameters: the rule, the element of the rule, the old value and the new value which should replace it.

\begin{table}[ht]
\caption{Atomic edit operations to modify ATL transformation programs, taken from~\cite{cuadrado2018-quickfix,DBLP:conf/models/VaraminyBahnemiry21}.}
\label{tab:edit-op}
\centering
\begin{tabular}{|l|c|}
\hline
\textbf{Target}                                                 & \textbf{Type}                                                                                  \\ \hline
Binding                                                & Creation                                                                              \\ \hline
Type of source pattern element                  & \multirow{4}{*}{\begin{tabular}[c]{@{}c@{}}Type \\ Modification\end{tabular}}         \\ \cline{1-1}
Type of target pattern element                      &                                                                                       \\ \cline{1-1}
 Type of collection                         &                                                                                       \\ \cline{1-1}
Type argument of operation            &                                                                                       \\ \hline
Navigation expression (binding RHS)                    & \multirow{2}{*}{\begin{tabular}[c]{@{}c@{}}Feature name \\ modification\end{tabular}} \\ \cline{1-1}
Target of binding (binding LHS)                        &                                                                                       \\ \hline
Predefined operation call  & \multirow{3}{*}{\begin{tabular}[c]{@{}c@{}}Operation \\ modification\end{tabular}}    \\ \cline{1-1}
Collection operation call             &                                                                                       \\ \cline{1-1}
Iterator call                  &                                                                                       \\ \hline
Binding                                                & Deletion                                                                              \\ \hline
\end{tabular}
\end{table}

\begin{figure}[t]
    \centering
    \includegraphics[width=\linewidth]{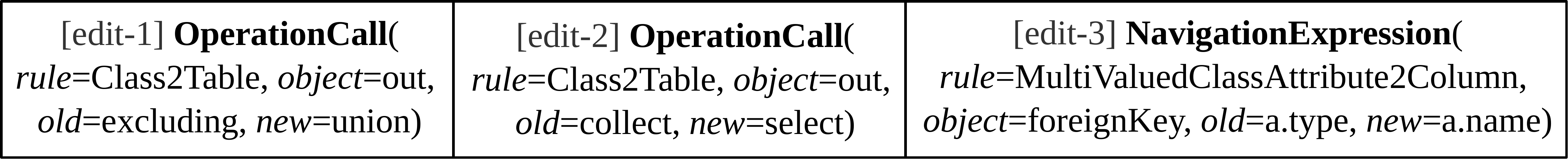}
    \caption{Example of a patch to repair the transformation program of Listing~\ref{fig:transformation_example} to conform to the behavior defined by the expected output model of Fig.~\ref{fig:comp_output_models}}
    \label{fig:optimal_patch}
\end{figure}

These edit operations can be used to compose patches to repair faulty ATL transformation programs.
Figure~\ref{fig:optimal_patch} shows an example of a patch composed of three edit operations applicable on the transformation program of Listing~\ref{lst:semanticerrors}.
The first operation replaces the operation call \texttt{excluding()} by \texttt{union()} in rule \textit{Class2Table} (line 7).
Similarly, the second operation replaces the operation call \texttt{collect()} by \texttt{select()} in the same rule (line 8).
The third operation changes a binding right-hand side in the rule \textit{MultiValuedClassAttribute2Column}: it replaces \texttt{a.type} by \texttt{a.name}.
This patch, when applied to Listing~\ref{lst:semanticerrors}, modifies the faulty transformation behavior: the obtained patched transformation now produces the expected output model.
This 3-edits patch is thus considered optimal to repair the transformation program with regards to the provided test cases.

\section{Problem statement}
\label{sec:problemstatement}

Designing patches to repair semantic errors is a difficult endeavor which requires an expertise in the transformation language, the meta-models and the transformation itself.
Input/output in test cases may reveal the presence of semantic errors, but do not provide a clear indication of what is causing the errors, nor the rules in which they may occur.
Detecting and fixing errors related to transformations' behavior is even more difficult because of the declarative nature of transformation languages such as ATL.
Moreover, gathering reusable knowledge about model transformation repair on which we could build automated or semi-automated approaches to assist experts in this task is tedious.
In fact, transformation programs are very dissimilar (notably because most of the program depends on the meta-models) and there are few available data about them. 
%\hs{Should we talk about the declarative nature of ATL transformation?}
In such situations, an alternative is to formulate the task as an \textit{optimization problem}, where the goal is to automatically find optimal solutions in the space of all possible solutions.

Formulating transformation repair as an optimization problem, an optimal solution represents a patch fixing the errors of the transformation program, and the space of solutions to be explored is thus equal to the set of all possible patches which could be applied on the faulty transformation.
However, this space cannot be explored exhaustively.
Indeed, for the large transformations with many errors, each error can potentially be repaired by choosing one or many edit operations, and each edit operation may involve any possible instance of elements in the input and output meta-models.  
Alternative methods are then necessary to efficiently explore this space.

\subsection{Evolutionary Algorithms}
Evolutionary algorithms (EAs) are search methods used to solve a wide range of optimization problems by efficiently exploring the search space.
Their search strategy is inspired by the evolutionary theory: 
EAs maintain a population of candidate solutions which undergo an evolution process through several generations.
At each generation, some solutions are mutated (i.e., we use an existing  solution to create a slightly different solution) and other solutions are bred (i.e., several existing solutions are recombined to create new solutions).
The newly created solutions along with the previous solutions are then evaluated and a fitness score is associated to each one of them, which reflects how good the solution is to solve the considered problem.
The solutions with the best scores have a higher probabilities to be retained in the population and to go through the next generation, while the others tend to be discarded.
By keeping the best solutions at each generation and using them to create new solutions, each new generation should have a population of solutions better suited to fix the problem than the previous one, until an optimal solution is finally found.

Population-based evolutionary algorithms have been studied to find patches to repair general purpose programs~\cite{forrest2009genetic,ding2019leveraging} and domain specific one such as ATL~\cite{DBLP:conf/models/VaraminyBahnemiry21,DBLP:journals/corr/abs-2012-07953}.
In this paper, we focus on the use of EAs for repairing model transformation programs.
Adapting a problem such as program repair to be solved with EAs revolves around three points: defining a solution representation, genetic operators and a fitness function.
In the rest of the section, we discuss these three points and illustrate them on the problem of repairing ATL transformation programs.

\paragraph{Solution representation}
In EAs, solutions are the central artifacts which are modified, evaluated and retained through generations.
Because this process is fully automated, choosing a way to represent solutions that ease their manipulation is essential for the approach to run smoothly.
Early EA-based approaches to repair programs used to consider a whole program as a solution: the population included different versions of the program to be repaired (usually in the form of ASTs) and evolved these programs until a correct version was found.
This could be costly in time and memory, and the evolution process was complex because it involved modifications on the AST.
A more convenient way to represent solutions in these cases is to consider patches in the form of sequences of edit operations as represented in Fig.~\ref{fig:optimal_patch}.
Sequences are easy to represent and manipulate, especially during the evolution phase, as discussed hereafter.

In the case of ATL transformation repair, a population would gather a set of patches being sequences of variable size of atomic edit operations, as presented in Section~\ref{sec:example-conformance}.

\paragraph{Genetic operators}
Genetic operators are at the core of the process of evolving solutions of each generation: they enable to obtain new candidate solutions based on the ones present in the population.
EAs usually rely on two types of genetic operators: \textit{mutation} and  \textit{crossover}.
The mutation operator takes one solution as input and outputs a slightly modified solution.
The crossover operator recombines two existing solutions (parents) to create two new solutions (children) composed of rearranged parts of their parents.
Usually, the operators are applied randomly until the population of solutions doubles in size.

\begin{figure}[ht]
    \centering
    \includegraphics[width=\linewidth]{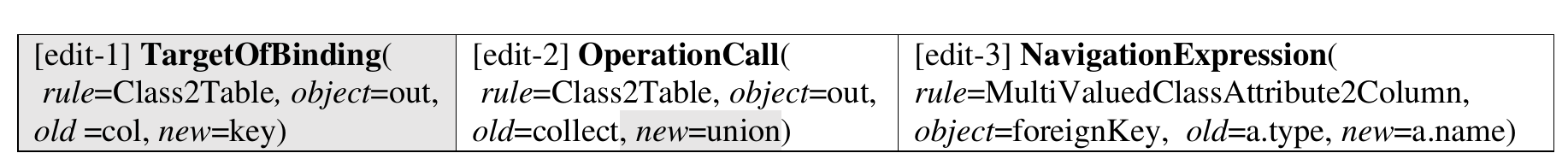}
    \caption{Examples of two mutations of the patch of Fig.~\ref{fig:optimal_patch}}
    \label{fig:mutation-op}
\end{figure}

In the case of evolving patches to repair faulty ATL transformations, the \textit{mutation operator} applies a mutation on one patch.
The considered mutations here are 1) adding an edit operation, 2) removing an edit operation and or 3) modifying an edit operation.
Fig.~\ref{fig:mutation-op} presents an example of two mutations applied on the patch of Fig.~\ref{fig:optimal_patch}.
The first mutation replaces \textit{[edit-1]} with another type of edit operation (target of binding) and the second mutation only modifies one parameter of \textit{[edit-2]}.

The \textit{crossover operator} takes two patches and outputs two new patches representing a recombination of the inputs.
In other words, it cuts the two sequences of edit operations in several parts (sub-sequences) and recombines them to create new sequences.
Representing solutions as sequences is thus convenient when performing crossover operations.
In this work, we used a single-point crossover operation, which separates patches in two parts and exchanges their right parts.

\begin{figure}[ht]
    \centering
    \includegraphics[width=.7\linewidth]{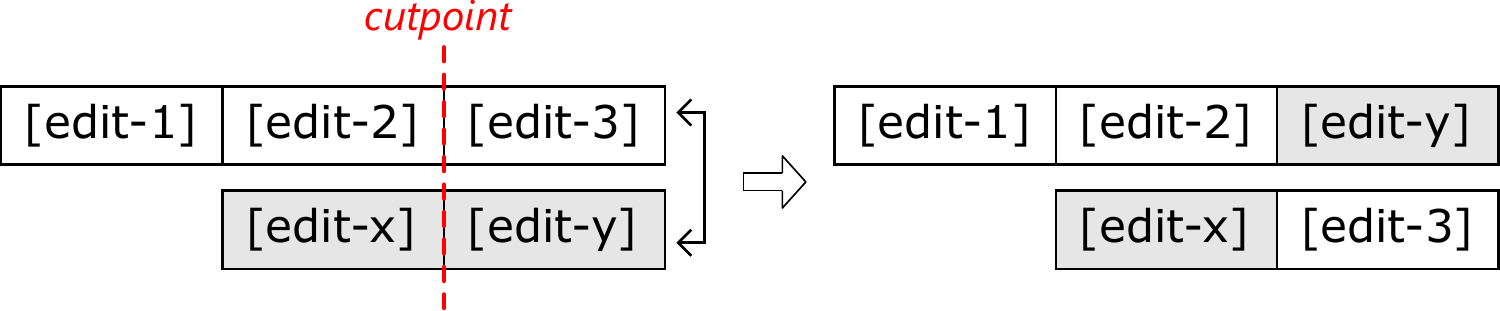}
    \caption{Examples of a single point crossover operation }
    \label{fig:crossover-op}
\end{figure}

Fig~\ref{fig:crossover-op} represents a single point crossover on the patch of Fig.~\ref{fig:optimal_patch} and another arbitrary patch of two edit operations.

\paragraph{Fitness function}
After the evolution phase, the fitness function is invoked on each solution to compute their fitness score.
This score should reflect how good is a solution to solve the problem, and is used to rank the solutions. 
This ranking is then used to select the better half of the population, and discard the solutions with poor fitness.

As explained previously, a way to detect the presence of semantic errors in ATL transformation program is by relying on input/output test cases.
If a patch, when applied to the faulty program, modifies the later such that it produces the expected output models for all input models, then the patched transformation is semantically correct with regard to the provided behavior specification, and the patch is thus considered optimal.
In this case, the fitness function could associate to each patch a score corresponding to the number of passing test cases.
At each generation, the fitness function would thus favor the patches passing the most test cases, until finding one passing them all.

\subsection{Issues with Convergence}

For a given problem, different fitness functions can be designed to achieve the same goal. Carefully designing the fitness function is essential and may impact both the approach's efficiency (time to converge toward an optimal solution) and efficacy (whether it converges towards optimal solution or not).
Indeed, the fitness score plays an central role in the search strategy of EAs, because selecting which solutions to retain or discard through the successive generations is what is \textit{guiding the search} by defining which parts of the solution space are explored or not.
Relying on test cases to assess the fitness of repair patches can can lead to convergence issues of EA repair approaches: we highlight two of them that we target in this paper.

Groups of similar solutions may have similar fitness scores.
Because EAs sift solutions with the highest fitness, it may promote groups of similar solutions if they have a high fitness score.
Mutations and crossovers, when applied on these solutions, will mostly produce similar solutions again, with high fitness as well.
Such groups may quickly overpower other solutions, leading to a loss of diversity in the population and a premature convergence toward a local optima.
This issue is known as \textit{single fitness peak}.
In the case where the fitness function relies on test cases, EAs will tend to promote patches correcting most of the errors.
We can end up in a situation where the population is mostly constituted of similar patches correcting the same errors and passing most of the test cases.
However, the other solutions that could target the remaining errors are quickly discarded in favor of these patches having a high score, and the necessary material to cover all errors and pass all tests is lost to their profits.
Sustaining a certain level of diversity within the population, i.e., ensuring that individuals are scattered in different regions of the search space, increases the chances to find good solutions efficiently.

\begin{figure}[ht]
	\centering
	\includegraphics[width=.6\linewidth]{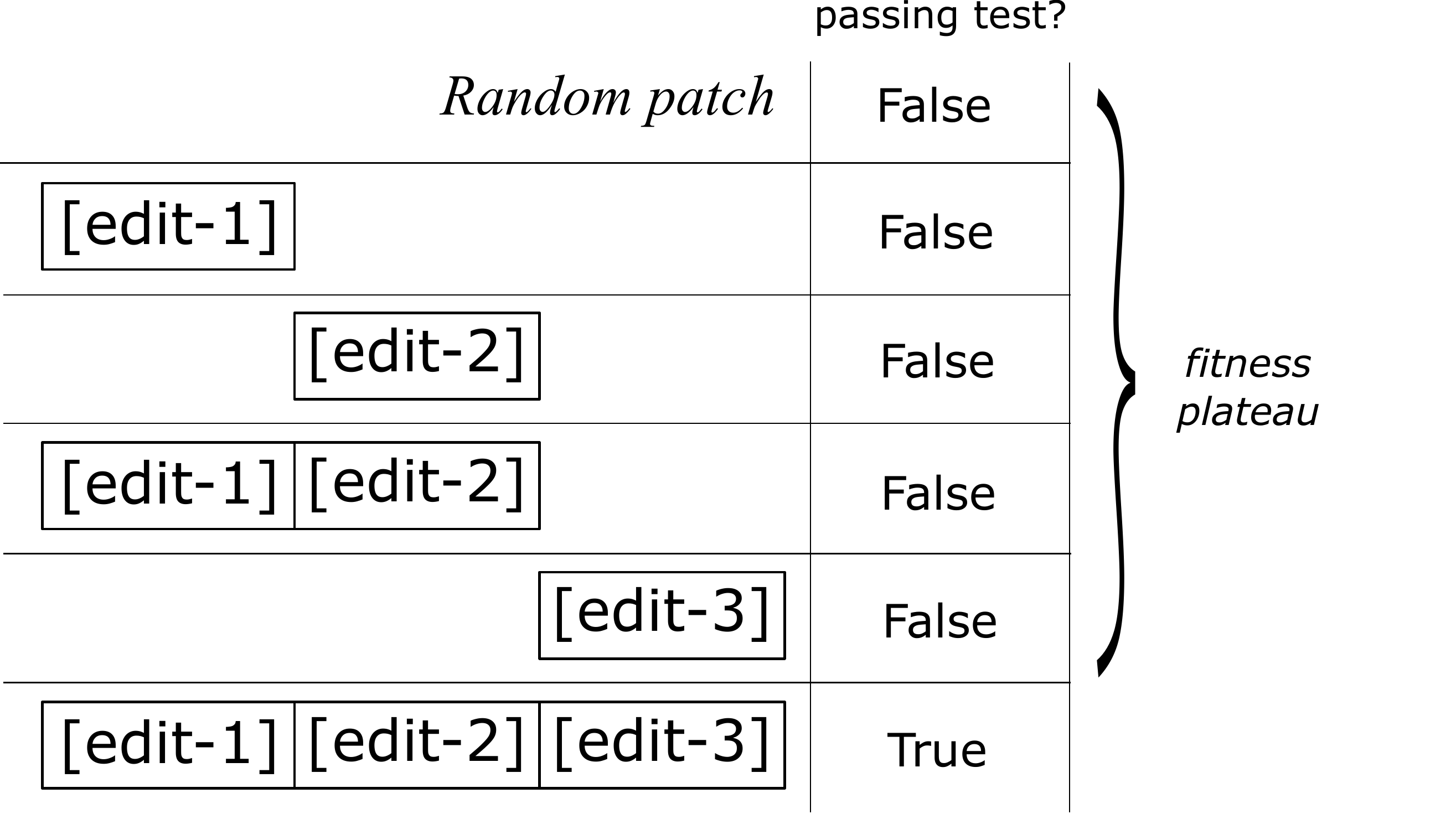}
	\caption{Example of fitness plateau caused by a fitness evaluation based on passing test cases}
	\label{fig:fp_ex}
\end{figure}

Using test cases to define fitness functions may lead to another issue hindering convergence: partial patches, i.e., correcting only a part of the defect, are associated with bad fitness score because test cases do not detect and reflect their value.
For instance, the patch presented in Fig.\ref{fig:optimal_patch} modifies the illustrative faulty transformation to pass the test case of Fig.\ref{fig:comp_output_models}.
However, sub-patches (or partial patches) such as \{edit-1, edit-2\} or \{edit-3\}, even though they correct part of the defect and are necessary to build the optimal patch, are not enough to pass the test, as illustrated in Figure~\ref{fig:fp_ex}.
These patches are thus indistinguishable from random patches which do not address at all the defects of the programs, and are discarded early in the process.
As a consequence, a lot of candidate solutions (partial or bad) have the same fitness score, thus creating \textit{fitness plateaus}, i.e., large parts of the fitness landscape where all solutions have the same fitness score even though they are different from one another, and even though some of them are partial solutions~\cite{de2018novel,ding2019leveraging}. 
This makes some parts of the search space difficult to explore, making it as good as random search because the fitness scores are the same and thus cannot properly guide the search. 

In a previous work~\cite{DBLP:conf/models/VaraminyBahnemiry21}, we used EAs to repair transformation programs by relying on test cases.
We obtained good results for faulty programs needing less than 3 edit operations to be repaired (i.e., with few errors): beyond this limit, our approach had trouble converging towards patches addressing all errors.
We analyzed in details the process of our approach for these cases, notably how candidate patches were selected or discarded through the generations, to understand why the approach was not effective anymore. 
We found that partial patches (partial solutions) were quickly discarded in the process due to fitness plateaus.
The more errors to correct, the bigger the size of the plateaus and the less effective the search for an optimal patch.

\section{Our approach}
\label{sec:ourapproach}

In this section, we propose a new approach for automated repair of semantic errors in transformation programs.
The goal of the proposed approach is to overcome the two aforementioned limitations faced by test-based EA approaches when repairing programs with several errors.
A hypothesis we make in this paper is that deliberately maintaining diversity in the population would not only help avoiding single fitness peak but also escaping fitness plateaus, hence increasing the effectiveness and efficiency of test-based EAs approaches.
The core of our approach is the use of several objectives to guide the search, including one to promote social diversity.
In what follows, we first  present the background related to multi-objective EAs. 
Then, we present the first objective, focusing on scoring patches depending on the provided test cases. 
We show how we use the expected output models of these test cases to retrieve more precise information regarding the program's errors and refine the fitness score. 
After that, we present the second objective designed to preserve diversity in the population. 
We discuss why we think that preserving diversity would help prevent both single fitness peak and fitness plateaus.
Finally, we present a third objective to prevent the patches to grow unnecessary large during the search, an issue known as \textit{bloating}.

\subsection{ Multi-Objective Evolutionary Algorithm}

Multi-objective optimization problems introduce the idea that candidate solutions may be evaluated based on several objectives, which may conflict with each other. 
Evolutionary algorithms are hence designed to find a set of near-optimal solutions, called non-dominated solutions (or Pareto front). 
A non-dominated
solution provides a suitable compromise
between all objectives without degrading any of them. Thus, non-dominated solutions are not comparable and can be considered equally good.
In this paper, we use NSGA-II~\cite{deb2000-nsga}, a well-known fast multi-objective genetic algorithm, that is suitable to the kind of problem we are solving~\cite{ali2020quality}.

\begin{figure}[ht]%nsga
	\centering
	\includegraphics[width=.6\linewidth]{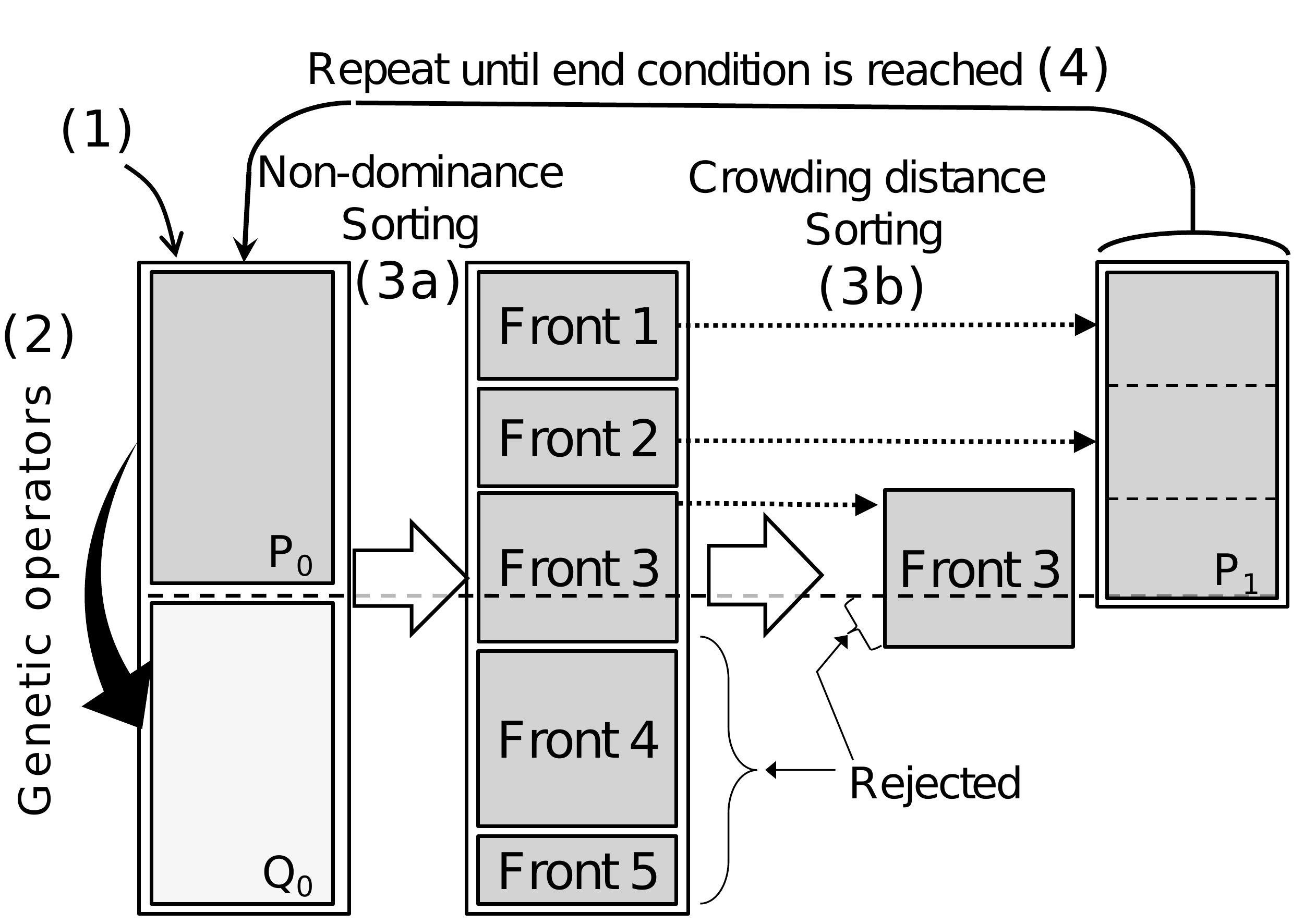}
	\caption{NSGA-II Algorithm~\cite{deb2000-nsga}}
	\label{fig:nsga2}
\end{figure}

Figure~\ref{fig:nsga2} presents the four main steps of NSGA-II.
The first step in NSGA-II is to create randomly a population $P_0$ of $N/2$ solutions (Fig.~\ref{fig:nsga2} (1)). 
Then, genetic operators are applied on the solutions of the population $P_0$ to create a child population $Q_0$ of the same size (2). 
Both populations are then merged into an initial population of size $N$. The populations are sorted into dominance fronts according to
the dominance principle (3a). 
A solution $s_1$ dominates a solution $s_2$ for a set of objectives $\{O_i\}$ if $\forall i, O_i(s_1) > O_i(s_2)$ and $\exists j | O_j (s_1) > O_j (s_2)$. 
The first front includes the non-dominated solutions. 
The second front contains the solutions that are dominated only by the solutions of the first front, and so on and so forth. 
The fronts are included in the parent population $P_1$ of the next generation following the dominance order until the size of $N/2$ is reached.
If this size coincides with part of a front, the solutions inside this front are sorted, to complete the population, according to a \textit{crowding distance} which favors diversity in the solutions~\cite{deb2000-nsga} (3b).
This process is repeated (4) until an optimal solution is found
in Pareto set or a stop criterion is reached, e.g., a number of iterations or one or more objectives greater than a certain threshold.

\subsection{Objective 1: Fixing as Many Errors as Possible}

The first objective addresses the main goal of the approach (i.e., repairing programs): it scores patches depending on their capability to fix errors in the faulty transformation programs.
As stated before, test cases are traditionally used to estimate the goodness of a patch: the more test cases pass, the better the patch.
In the case of ATL transformation programs, test cases are pairs of input/output models: provided with the input models, a correct transformation should output the expected models.
To assess a patch, first the sequence of edit operations is applied on the faulty transformation to obtain a patched transformation.
Then, the input model of the test case is given to the patched transformation to produce an output model: if the obtained model is equivalent to the expected output, then the test case passes. 
If the obtained model is different from the expected one, the test fails.

Failing test cases do not usually provide information regarding why they fail, or how close they were to pass.
However, working with test cases based on model comparison gives us the opportunity to refine the fitness score by considering the differences between the two output models. 
The idea is that even if a patch does not correct all the errors and does not pass all the tests, a partial solution should lead to less discrepancies between the output models and the expected ones compared to a random solution.

For this objective, we compare the output models generated by the patched transformation with the expected ones provided by the test cases to compute the number of differences between these models.
We rely on EMFCompare~\cite{brun2008model}, a tool which, given two models, output a list of differences between them, in a similar manner than the differences presented in Fig.~\ref{fig:comp_output_models}.
The higher the number of differences, the less fit the patch is considered, and it thus receive a bad score.
An optimal patch is a patch for which the number of differences is zero.
When such a patch is found, the process stops.

\begin{figure}[ht]
    \centering
    \includegraphics[width=.6\linewidth]{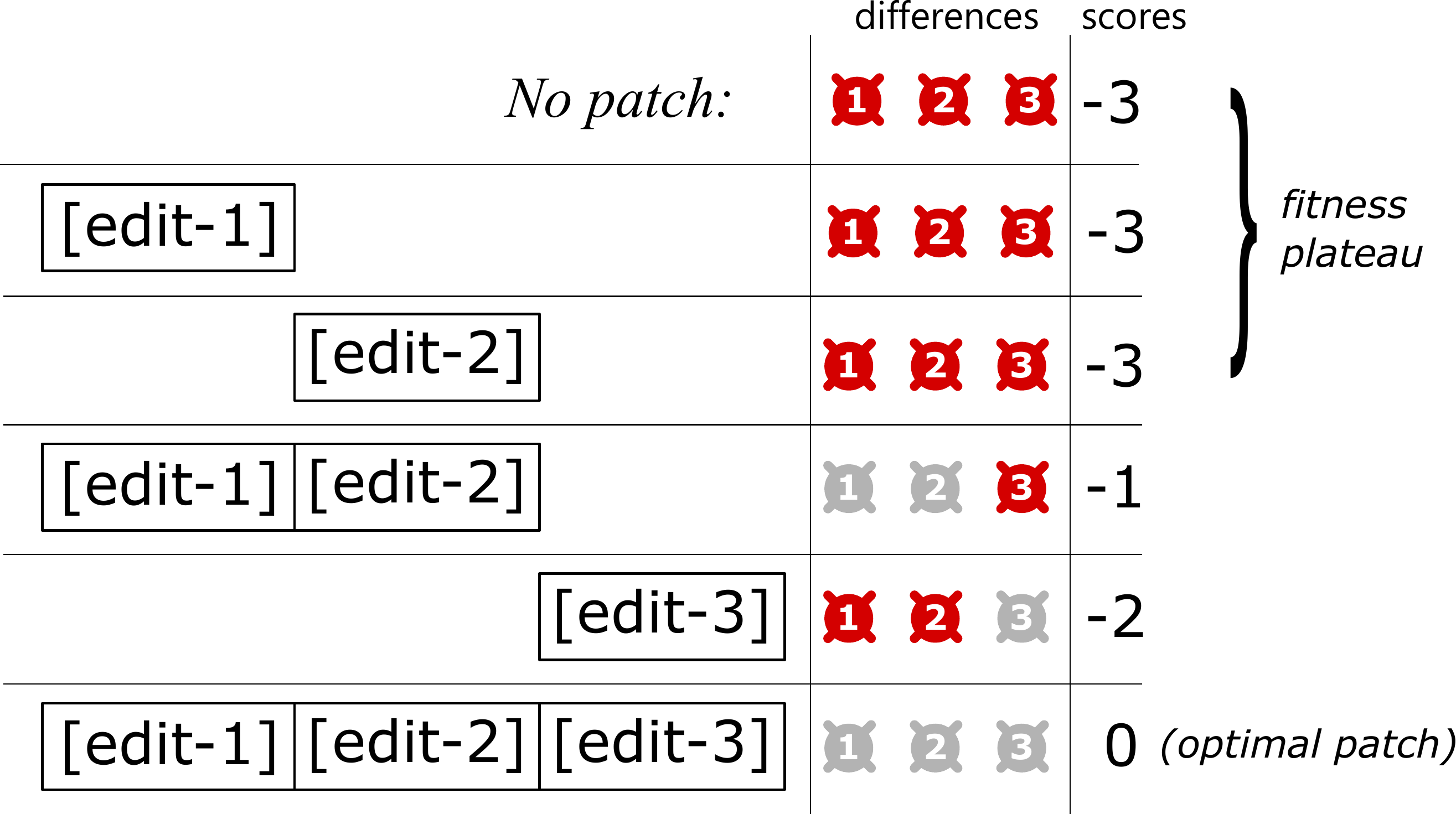}
    \caption{Assessing patch fitness depending on model differences}
    \label{fig:patch_fitness}
\end{figure}

The patch presented in Fig.~\ref{fig:optimal_patch} is optimal because it produces the expected output model:  it fixes the 3 differences indicated in Fig.~\ref{fig:comp_output_models}.
Non-optimal patches could either (a) introduce new differences, (b) do not change the output models or (c) correct some differences but not all.
Figure~\ref{fig:patch_fitness} shows an example of scores given to partial patches inspired from the optimal patch of Fig.~\ref{fig:optimal_patch}.
The unmodified faulty transformation (no patch) produces an output model with 3 differences.
If we consider a patch composed of the two first edit operations of Fig~\ref{fig:optimal_patch}, it fixes differences 1 and 2, but not 3.
A patch composed only of the third operation \textit{[edit-3]} fixes the difference 3 but not the differences 1 and 2.
This patch can be thus not considered as good as the previous one, because it fixes one less difference.
However, it is still better than no patch at all.
To properly assess the fitness of patches and compare patches, it is best to consider several examples to approximate the expected behavior of a transformation.

Even if this score is more fine-grained than relying on the number of passing test cases, it is not enough to prevent fitness plateaus~\cite{DBLP:conf/models/VaraminyBahnemiry21}.
It appears that in cases where several errors needed to be corrected in one program, many of them needed to be corrected at the same time to see an improvement in the output models. 
If two errors disturb similar parts of the output models, correcting only one of them would not improve the output models, thus both of them need to be corrected at the same time to notice any improvement, hence hindering the detection of partial solutions, as shown in Fig.~\ref{fig:patch_fitness}. [edit-1] and [edit-2] both modify the same binding (lines 7-8) in the rule Class2Table.
To notice an improvement in the fitness scores, they both need to be present in the patch, otherwise, the score is as good as for no patch at all.
However, the fitness plateau is smaller than the one shown in Fig.~\ref{fig:fp_ex}, when only the numbers of passing test cases were used to assess the fitness of a patch.

\subsection{Objective 2: Preserving Semantic Diversity}

Our second objective focuses on promoting the diversity in the population.

The literature recognizes two types of diversity.
The first one, called \textit{genotypic} or \textit{syntactic} diversity, distinguishes individuals based on their structure.
In our case, syntactic diversity would promote patches of variable size and using dissimilar edit operations.
The second type of diversity is called \textit{phenotypic} or \textit{semantic}. 
This time, it distinguishes individuals based on their behaviors without considering their structure.
Patches having similar size and edit operations but modifying the transformation programs such that they result in different output models would then be considered semantically diverse. 
When targeting semantic errors in programs, maintaining diversity in programs' behaviors is highly relevant.
On the other hand, understanding the impact of syntactic diversity on the programs' behaviors is  quite complex~\cite{mcphee1999analysis}.
We thus focus on semantic diversity, which is also known to be more efficient to prevent single fitness peak~\cite{batot2018injecting,vanneschi2014survey}.

\textit{Social} semantic diversity is particularly interesting in our case.
The aim of a social diversity measure is to assess a candidate solution not only by examining the solution alone, but also by considering the solution as a part of the population. 
When repairing transformation programs, a social diversity measure would consider that the value of a patch should not be restricted to the number of errors it corrects, but should also consider its capability to address errors which are infrequently covered by the other patches of the population. 
%Thus, the value of a candidate does not only depend on its capacity to address the problem at stake, but also consider the capacity of the other candidates to solve it. 
In other words, it aims at assessing the value a candidate patch brings to the entire population.

Batot et. al~\cite{batot2018injecting} proposed a social diversity measure giving higher scores to solutions which pass test cases frequently failed by the other solutions.
A solution passing numerous test cases that the majority of the population also pass will receive a lower score than a solution passing less test cases but which are failed by a majority of the population.
They show that considering this measure to score solutions allows to reduce single fitness peak when using test cases conformance to guide the search.

In this paper, we propose a social diversity measure relying, not on the  number of passing test cases, but on the differences between the obtained output models and the expected ones.
Because these differences give information about what part of the output models differ from what is expected, we are able to estimate which parts of the output models are impacted by each patch.
We use this information to give higher scores to patches modifying parts of the output models which are less covered by the other patches of the population.

As discussed previously, correcting programs with many errors increases the chances to have several errors impacting the same parts of the output models.
These errors need to be fixed at the same time to notice a difference in the output model, and thus an improvement in the fitness score.
We think that bringing social diversity in our fitness function will help maintain a population of patches addressing different parts of the output models, thus increasing the chances to escape fitness plateaus caused by errors interactions.
Moreover, using a social diversity measure as an objective would refine the fitness score by adding a new level of granularity, thus helping reduce the size of the plateaus.

\subsection{Objective 3: Controlling the Size of the Generated Patches} 

Bloating is a known issue in EAs where the solutions considered during a run grow in size and become larger than necessary to represent good solutions. 
This is unpleasing because it slows down the search by increasing manipulation and evaluation time, and find good solutions which are unnecessary large and complex.
In multi-objective EAs, dedicating an objective to give better scores to solutions of small size (Parsimony Pressure) appeared to be effective to prevent bloating.
Thus, we use a third objective,  which represents the number of operations in the patch, to favor patches of small size to avoid to generate candidate patches using too many edit operations.

\section{Experiments}
\label{sec:experiments}
We implemented our approach in a tool, called Automatix, and performed an empirical
evaluation\footnote{All experiment data and code are available at \url{https://github.com/zahravaraminy}}
This section reports on the evaluation of the impact of our multi-objective approach using social diversity on correcting several semantic errors.
We perform our evaluation on four existing third-party transformations, \textit{Class2Table}, \textit{PNML2PN}, \textit{Bibtex2Docbook} and \textit{UML2ER}. 
We formulate the following research questions: 
 
\textbf{RQ1:} What is the impact of social diversity on the effectiveness of the approach (i.e., finding a patch correcting all the errors)?

\textbf{RQ2:} What is the impact of social diversity on the efficiency of the approach (i.e., the convergence time)?
 
\textbf{RQ3:} What is the impact of social diversity on the type of errors which are corrected?

\subsection{Dataset}
We performed our evaluation on existing faulty transformation programs from the literature.
In  previous work~\cite{DBLP:conf/models/VaraminyBahnemiry21}, we utilized 13 faulty versions of the Class2Rel transformation, and 18 faulty versions of the PNML2PN transformation. 
We reused the 31 faulty transformations studied in this paper.

Then, we complete this dataset with transformations having three errors or more to assess the impact of social diversity in these cases.
Guerra et al.~\cite{guerra_towards_2019} introduced an approach for mutation testing in ATL transformations. 
Mutation testing process needs mutants coming from distinct error categories.

We retrieved the UML2ER mutants and the Bibtex2Docbook mutants from their paper. 
We tested each mutant with the AnAtlyser tool~\cite{cuadrado2018-anatlyser}, which finds a wide range of syntactic errors (including type errors) in ATL transformations using static analysis.
We only select mutants containing semantic errors and we discarded the mutants with syntactic errors. 
Out of the 800/354 mutants for Bibtex2Docbook/UML2ER studied in their paper, 101/48 of them were syntactically correct but presented semantic  discrepancies  with  the  original  transformations.
However, these mutants only have one semantic error. 
We reused the approach presented in~\cite{DBLP:journals/corr/abs-2012-07953} to merge several mutants with one error to obtain mutants with several errors.
We applied this approach on UML2ER and Bibtex2Docbook mutants to create mutants with multiple semantic errors. 
Previous work~\cite{DBLP:journals/corr/abs-2012-07953} showed that multi-objective GP face convergence issues to repair faulty transformations having 3 or more errors.
To study the impact of social diversity on higher numbers of errors, we thus created 4 sets with respectively 2 to 5 mutants and then merged them in each set to form 4 faulty transformations with 2 to 5 semantic errors. 
We ran this creation process 5 times for both UML2ER and Bibtex2Docbook mutants. 
In the end, we acquired 20 faulty versions of each transformation
(5 * 4 faulty transformations having 2 to 5 errors). Table~\ref{tab:transfo} presents information characterizing the 4 transformations and their input/output metamodels.

\begin{table}[ht]

\caption{Transformations used in the evaluation. Cells with two values represent  input/output metamodels}
\label{tab:transfo}
\begin{tabular}{r|cccc}
                                                                    & Classe2Table & PNML2PN & Bibtex2Docbook & UML2ER \\
                                       \hline
LoC                                                                 & 136          & 91      & 232 & 79      \\
Rules                                                               & 8            & 5       & 9 & 8       \\
Helpers                                                             & 4            & 0       & 4 & 0         \\
\hline
Classes                                                             & 6/5          & 13/9    & 21/8   & 4/8\\
Attributes                                                          & 3/1          & 4/3     & 9/2  & 87/2 \\
Associations                                                        & 11/8         & 28/20   & 21/9 & 7/10  \\
\begin{tabular}[c]{@{}c@{}}Inheritance \\ associations\end{tabular} & 5/3          & 14/8    & 18/4  & 3/7
\end{tabular}
\end{table}

We studied on the size (in terms of number of rules) of the 106 ATL transformations from the ATL Zoo and found out that a transformation has an average of 11 rules, with Q1 = 5, Q3 = 12 and the median being 9.
The four selected transformations of our evaluation are thus representative of transformations found in the ATL Zoo.

To identify semantic error types in faulty transformations, we determine how many atomic modifications need to be performed to correct it: the type of elements of the faulty transformation that should be modified determine the semantic error type.
Types of semantic errors are thus strongly related to the edit operations used in this approach.
We identified 9 kinds of elements that could be modified by an atomic edit operation: the types of input/output patterns, the operation calls and their arguments, the types of collections, the properties of input/output object, the bindings (missing bindings and extra bindings).
Table~\ref{tab:errors} presents the 9 different classes of semantic errors, as well as their occurrences in the faulty transformations used in the experiments. We can see that each error type is well represented in our dataset.

\begin{table}[ht]
\caption{Classes of semantic errors that can be found in ATL transformation programs.}
\label{tab:errors}
\centering
\begin{tabular}{c|lc}
\textbf{Id} & \textbf{Type of semantic errors} & \textbf{Occurrences}\\
\hline
TOP         & Wrong type of output pattern & 32    \\
TIP         & Wrong type of input pattern & 13     \\
OP          & Wrong operation call  & 22           \\
TA          & Wrong type argument & 19             \\
CT          & Wrong collection type & 9           \\
BL          & Wrong property in binding LHS & 29    \\
BR          & Wrong property in binding RHS & 30    \\
MB          & Missing binding & 29                 \\
EB          & Extra binding         & 21          
\end{tabular}
\end{table}

\subsection{Process}
In this experiment, we aim at testing social diversity with two configurations (as a crowding distance and as an objective) separately, because we think that they can help assessing the impact of social diversity on convergence from two different perspectives.
Using a social diversity measure as a crowding distance will help hamper a loss of diversity without altering the fitness function. 
Thus, it would help understand how diversity in the population impact the resolution of problems whose fitness landscapes contain large plateaus, and thus if social diversity can help escape such plateaus.
Using a social diversity measure as an objective would refine the fitness score by adding a new level of granularity, thus helping reduce the size of the plateaus. 
This time, we could see how social diversity impact the fitness landscape and if it makes it easier to explore.

We thus adapt our approach to run with three different configurations: without social diversity, with social diversity as a crowding distance and with social diversity as an objective.
We run our approach on all faulty transformations (71 in total) with the three configurations to compare the results.
We  set a  maximum  number of generations  to  50 000.
If an optimal patch, which fixes all the semantic errors, is found before attaining the 50 000 generation, the program stops and the number of generations needed to find the patch is preserved. 
If  no  optimal  patch  is  found at the end of the 50 000 generations,  we  retain  the  best patch  found (i.e., the one with the best fitness score) at  the  end  of  the  last  generation. 
Because EAs are probabilistic approaches, we run our process 5 times on each faulty transformation and for each configuration to be able to compute averages.
We thus run the 71 distinct faulty transformations 5 times, for a total of 355 runs for each configuration.

To answer RQ1, we compare the effectiveness of each configuration, i.e., the number of time a run can find an optimal patch.
To answer RQ2, we compare the efficiency of each configuration, i.e., the number of generations necessary for a run.
To answer RQ3, we applied the obtained best patches on the faulty transformations, and we manually compared the patched transformations with their correct versions to identify the number of remaining errors (if the patch was not optimal) and their types.

\subsection{Results}

\paragraph{RQ1: What is the impact of social diversity on the effectiveness of the approach (i.e., finding a
patch correcting all the errors)?}

\begin{figure*}[ht]
    \centering
    \includegraphics[width=\linewidth]{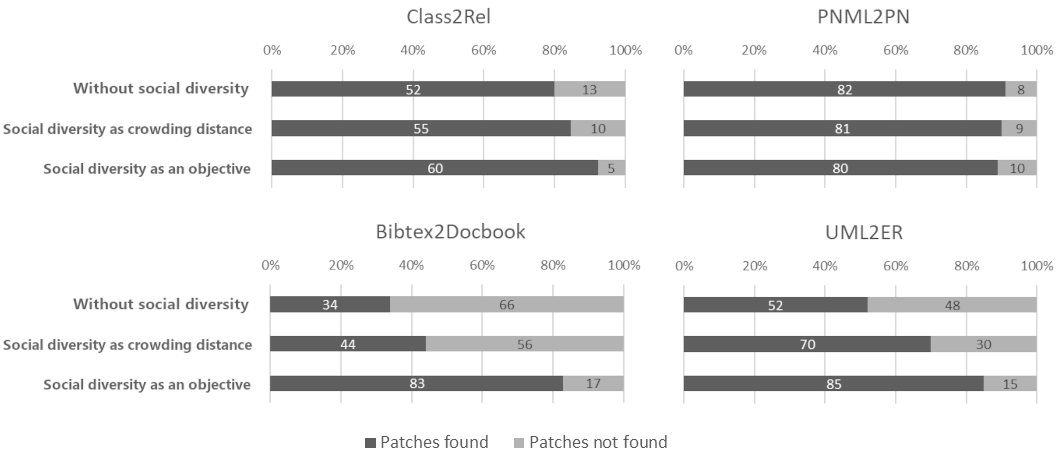}
    \caption{Percentage of runs finding an optimal patch (RQ1)}
    \label{fig:RQ1}
\end{figure*}

The percentages of runs that find an optimal patch for all four transformation programs are shown in Fig~\ref{fig:RQ1}.
The results shows an improvement in finding optimal patches in configurations using social diversity in three problems out of four:
\textit{Class2Rel}, \textit{Bibtex2DocBook} and 
\textit{UML2ER}.

Class2Rel and PNML2PN mostly include transformations with one or two errors.
As we have seen before, the patch generation approach without social diversity already worked effectively in these cases, which explain why social diversity does not introduce huge improvements.
Note that among the transformations studied in our previous work~\cite{DBLP:conf/models/VaraminyBahnemiry21}, Class2Rel regrouped the largest and more complex ones, which were the most difficult to handle with the approach without diversity.
Even if the improvement is not important, it is still noticeable that injecting social diversity helped increase the effectiveness of these difficult cases.
PNML2PN is the only problem in which social diversity does not increase the effectiveness of the initial approach.
However, the percentages are so close that they are not really significant: we cannot conclude that social diversity reduce the effectiveness.
PNML2PN is less complex than Class2Rel and contains few transformations with more than 2 errors.
The configuration without social diversity already gives very good results on this case (near 90\% of the runs found an optimal patch): social diversity does not bring improvement in easy cases.

For \textit{Bibtex2DocBook} and \textit{UML2ER}, we noticed sizable improvements for finding optimal patches when injecting social diversity in the process. 
In both cases, social diversity as an objective yields better results than as a crowding distance.
Because these two cases mostly contain transformations with 3 errors or more, we expect their fitness landscape to contain more plateaus than the ones of Class2Rel and PNML2PN.
The results of social diversity as a crowding distance suggests that diversity indeed helps escape these plateaus in certain cases, improving the effectiveness from 34\% to 44\% for Bibtex2DocBook, and from  57\%  to 70\% for UML2ER.
But considering social diversity as an objective (resulting in reducing the size of plateaus) give even better results, attaining an effectiveness of 83\% and 85\% for Bibtex2DocBook and UML2ER, respectively.

We can conclude that using social diversity both as crowding distance and as objective improves the correcting larger number of errors at the same time.

\begin{figure*}[ht]
    \centering
    \includegraphics[width=\linewidth]{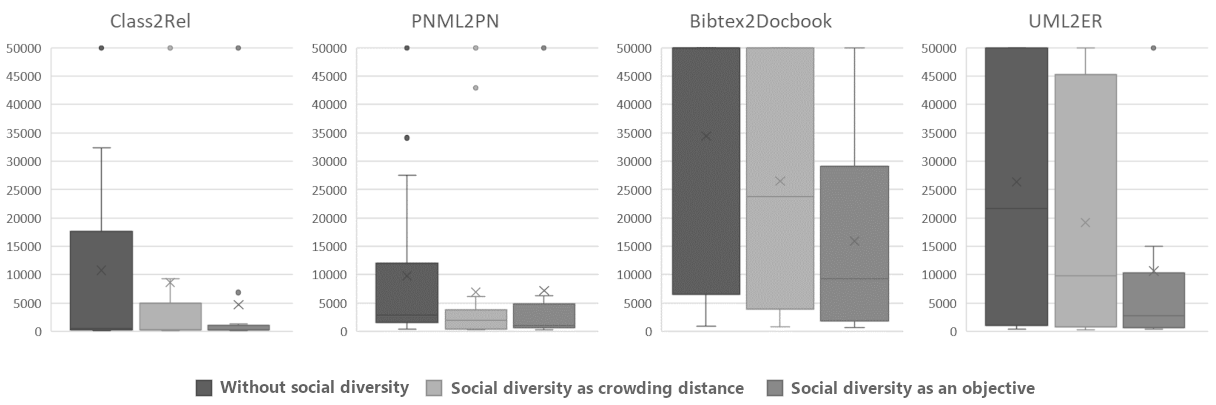}
    \caption{Number of generations (convergence time) to find a solution }
    \label{fig:RQ2}
\end{figure*}

\paragraph{RQ2: What is the impact of social diversity on the efficiency of the approach (i.e., the convergence
time)?}
Figure.~\ref{fig:RQ2} shows the average number of generations to obtain a solution for the four studied transformations, and depending on the three configurations.

Overall, the average number of generations required to find a good patch when injecting social diversity in the process is smaller for all four transformation programs. 
In RQ1, we saw that social diversity do not substantially increase the effectiveness of the approach for Class2Rel and PNML2PN, which gather transformations with small numbers of errors. 
However, Fig.~\ref{fig:RQ2} shows that social diversity improves its efficiency.
Here again, social diversity as an objective give better results than as a crowding distance for Class2Rel.
For PNML2PN, social diversity as an objective or as crowding distance gives similar results, but they are both better than the initial configuration without social diversity.

For Bibtex2Docbook and UML2ER, even though the convergence time is better with both configurations including a social diversity measure, the one adding social diversity as an objective brings a higher improvement than the one using social diversity as a crowding distance.
In fact, for these transformations with many errors, injecting social diversity through the crowding distance is more effective than the initial approach but the differences are not that important.
This suggests that injecting diversity without altering the fitness function increases the chances to find optimal patches (see RQ1), but the exploration is still difficult and the convergence takes time.
Reducing the plateaus' size by introducing the diversity measure as an additional objective, however, seems to ease the exploration process, leading to a fastest convergence and a better effectiveness.

Thus, we conclude that using social diversity helps the approach find the optimal solutions faster.

\subsection{RQ3: What is the impact of social diversity on the type of errors which are corrected?}

\begin{figure}[ht]
    \centering
    \includegraphics[width=\linewidth]{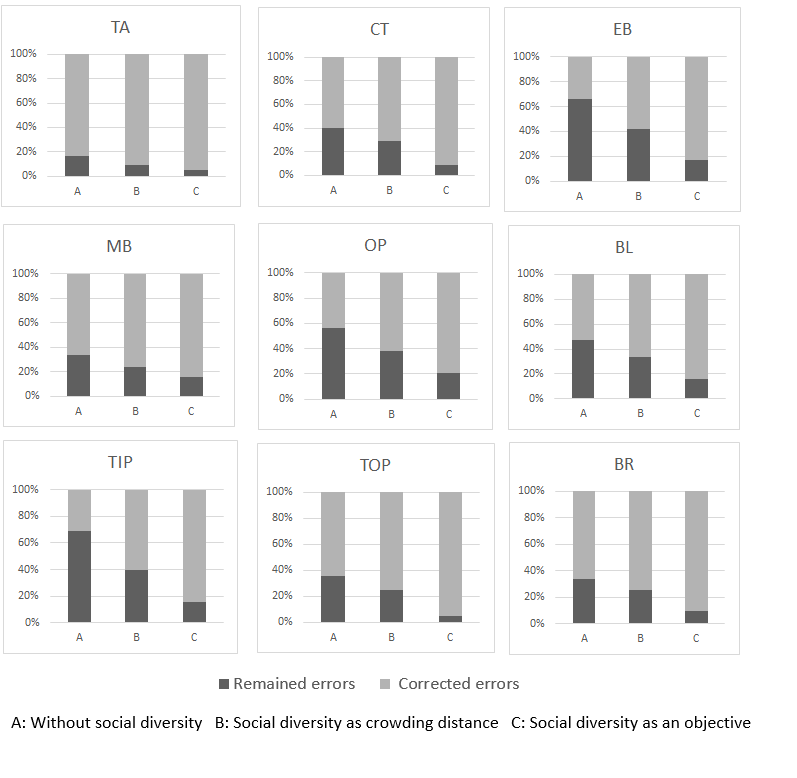}
    \caption{Percentage of corrected/remained errors for each type of semantic error in faulty transformations
    }
    \label{fig:RQ3}
\end{figure}

To answer RQ3, we first retrieve the number and type of errors present in all faulty transformations.
For each type of semantic error, we computed their occurrences in the studied faulty transformations. 
Since we run each faulty transformation 5 times for each configuration in our approach, we multiply the total number of errors 5 times to correctly calculate the ratio of corrected/remaining errors. 
Finally, we counted the total number of errors that are corrected/not corrected by the best patch found at each run.
We repeated this process for the three configurations. 
At the end, we obtained, for each error type, the total number of their occurrences in the faulty transformations and the percentage of corrected errors for each configuration, as shown in Fig.~\ref{fig:RQ3}.

We can see that injecting social diversity, both as a crowding distance and as an objective, increases the correction rate for all types of errors. 
For example, the  correction rate of semantic errors related to a wrong type argument (\textit{TA}) increased from 83.15\% (without SD) to 90.53\% (social diversity as a crowding distance) and to 94.74\% (social diversity as objective). 
Errors of type EB, OP and TIP are difficult to repair without social diversity: more than 50\% of them remain after applying the best patch. 
Considering a social diversity measure in the fitness function allows to decrease this percentage to 20\% or less for these three cases.
Here again, social diversity as an objective provides better improvement than social diversity as a crowding distance.

Even if social diversity improves the correction rates of all types of errors, some of them remain more difficult to fix than other. 
Those include the three types which were the most difficult to handle without social diversity (EB, OP, TIP).
We performed a behavior analysis of our automated approach, especially on the candidate patches which are discarded or kept at each iteration, to understand why some errors remain more difficult to correct than the others.
We observed that combinations of these types of errors are more likely to cause interaction, i.e.,  impact the same parts of the output models and need to be fixed at the same time to see a improvement in the fitness score.

This evaluation shows that social diversity can help overcome the limitations caused by fitness plateaus, which occur when trying to find complex patches (i.e., correcting several errors) while guiding the search with test cases. 
We showed that in our case, injecting social diversity in the population helps improve the effectiveness of the approach for repairing more than 2 errors.
The convergence time is also reduced but remain high, suggesting that the exploration is still difficult.
We also showed that refining the fitness function by adding social diversity as an objective improve both the effectiveness and the efficiency of the approach.
It creates a smoother fitness landscape, more suited for the exploration process. 
Finally, this evaluation highlighted that some types of errors are more difficult to repair than other, because they are more likely to form fitness plateaus when combined with each other.

\subsection{Threat to Validity}

There are some threats to validity in our approach as follows.
A first and main threat to validity is the input/output model examples to evaluate the behavior of a transformation, which may not cover all types of semantic errors in a transformation. This causes the incorrect transformations produce expected output models. We used four different input/output examples to overcome this threat. Second threat to validity is that we are fixing ATL transformation rules not helpers. We think that if we define the edit operations, which can modify helpers in ATL transformations, then we can use them in our approach and also fix helpers.
Another limitation of our work is that 
we tested our approach only with the Atlas Transformation Language (ATL) but
we believe that our approach can be generalized with other transformation languages using specific version of edit operations related to the targeted language. The semantic errors in faulty transformations used in the evaluation originated from mutants and not actually introduced by developers. We used this external data set since it is independent from our project and it covers a large spectrum of semantic errors.

\section{Related Work}
\label{sec:relatedwork}
The work presented in this paper intersects three research areas: program repair in general,  repairing  transformation programs and social diversity. In the following subsections, we discuss this three areas.

\subsection{Model repair}

Ben Fadhel et al.~\cite{Fadhel-detection} use
a search-based algorithm to express high-level model changes in terms of refactorings. 
Their approach takes a list of possible refactorings, an initial model and its revised version, and searches for a sequence of refactorings characterizing the changes made to obtain the revised model. 
After applying the sequence of refactorings on the initial model, the obtained  model should be as close as possible as the provided revised model. 
Their approach finds a sequence of edit operations based on model differences but our method applies on transformations. 

Puissant et al.~\cite{puissant2015resolving} proposed an approach to resolve model
inconsistencies. 
They use automated planning to generate one or more resolution plans to repair one error.
A change-preserving model repair approach is proposed by Taentzer et al.~\cite{Taentzer-change-preserving}, based on the theory of graph transformation. They consider the edit operations history to identify the inconsistent changes in a model, and complete them with number of possible repair actions to restore consistency.
A rule-based repair of EMF models  with user intervention
is proposed by Nassar et al.~\cite{Nassar-rulebased}. 
Their approach repair models 
in a specific context but the efficiency of evolutionary algorithms, in which we used in our approach, is independent from the context. In~\cite {Kretschmer-validationtree}, Kretschmer et al. present an automated approach to explore the space of possible repair values using validation trees to repair model inconsistencies. In comparison, in our approach 
we explore the space of possible patches using evolutionary algorithms, which is based on random choices and genetic operators, and leads to more diverse solutions.
Bariga et al.~\cite{DBLP:conf/models/BarrigaMPRHI20} presented an automatic model repair method which uses reinforcement learning, in which used Markov Decision Process (MDP) and Q-learning algorithm, to repair broken models. 
Their goal is to generate sequences of edit operations to apply on the whole model, and not just specific errors.

\subsection{Transformation repair}

Troya et al.~\cite{troya2018-spectrumbased} proposed a Spectrum-Based Fault Localization  technique, which uses test cases to find the probability of transformation rules being faulty. Oakes et al.~\cite{OakesTLW182018} presented an approach to statically
verify the declarative subset of ATL model transformations.
They translated 
the transformation into DSLTrans and used a symbolic-execution approach to 
produce representations of all possible executions to the transformation. They verify pre-/post-condition contracts on
these representations to verify the transformation.
These two works focus on detecting faulty rules in transformation programs and do not repair the faulty rules.
Burgueño et al.~\cite{burgueno2015-staticlocalization} presented  a static approach to check the correctness of transformation rules using
 matching functions, which used metamodel footprints to
automatically generate the alignments between implementations and specifications. Cuadrado et al.~\cite{cuadrado2014-uncovering}  presented a combined method
using
a static analyzer and a constraint solver to detect errors in model transformations. They produced a witness model using constraint solving to make the transformation to execute the erroneous statement. These approaches could find the faulty rules in model transformation, but they cannot fix transformation errors.
Cuadrado et al.~\cite{cuadrado2018-quickfix}
proposed a tool, Quick fix, to repair syntactic errors in ATL transformations using a static 
analyzer proposed in ~\cite{cuadrado2014-uncovering}.
Their approach needs a user interaction to select a suitable repair. On the other hand, our approach generates a candidate patch automatically. 
In a previous work~\cite{DBLP:journals/corr/abs-2012-07953}, we relied on the static analyzer of~\cite{cuadrado2014-uncovering} to automatically generate patches addressing syntactic errors in transformation programs.
An impressive work is presented by Kessentini et al. in ~\cite{kessentini2018-MM-MT-Coevolution} that implemented an evolutionary algorithm to modify a model transformation to conform to new versions of the metamodels. 
Their approach aims to adapt models to the new version of metamodels syntactically but not semantically. 
Rodriguez et al.
  ~\cite{rodriguez2021suggesting}
  proposed the Model Transformation TEst
Specification (MoTES) approach to repair
 transformations for rule-based languages.

Their approach is based on a metric-based test oracle and they used  input/output models to
mark input/output pattern relationships as true positive, true negative, false positive or false negative.
In our approach, we used input/output models as a measure of diversity to choose candidate patches which are less similar to the others for next generation.

\subsection{Social diversity}

Soto \cite{Soto2019} proposed a study of patch diversity as a means to increase the quality of generated patches through patch consolidation. 
Their approach focuses on improving patch quality for general program repair.
Ding et al. \cite{ding2019leveraging}
used a search-based technique for program repair, which is successful when it produces short repairs. The fitness function  relies on test
cases, which are not enough to determine partially correct solutions
and lead to a fitness plateaus.
They proposed a novel fitness function  using learned invariants over intermediate behavior. Their approach improved semantic diversity and fitness but not repair performance.   This approach is similar to ours in the sense that they used the semantic diversity to optimize the fitness function. However, They used invariant-based semantic diversity but we used 
social diversity in different way.
Their method applies on programming languages but ours applies on transformation languages.

Vanneschi et al.
 \cite{vanneschi2014survey}
divided semantic-aware methods into three categories. \textit{Diversity methods},
that work with diversity, mostly at the population level \cite{koza1992programming}.
\textit{Indirect semantic methods}, that act on the syntax of the individuals and depend on criteria to indirectly promote a semantic behavior \cite{o1998fitness}
\cite{galvan2013locality}
\cite{galvan2011defining}
\cite{krawiec2009approximating}
. \textit{Direct semantic methods}, that act directly on the semantics of the individuals by using precise genetic operators
\cite{moraglio2004topological}. All these approaches help improving the power of genetic programming.
Batot  et  al.  \cite{batot2018injecting} proposed injecting social diversity in multi-objective
genetic programming to learn model well-formedness rules from examples and tackle the bloating and single fitness peak limitations. They presented an improvement in population's social diversity  that was performed during the evolutionary computation and lead to efficient search strategy and convergence. They implemented the social semantic diversity in NSGA-II algorithm both as crowding distance and as an objective. 
The difference with our work are that we aim at fixing semantic errors in ATL transformations not learning model well-formedness rules from examples. 
Interestingly, they obtained better results when injecting social diversity in the crowding distance than as an additional objective.
This could be explained by the fact that we target two different issues: they try to limit the loss of diversity to prevent single fitness peak while we try to overcome the issues caused by fitness plateaus.

\section{Conclusion and Future Work}
\label{sec:conclusion}
In this paper, we presented a novel automated approach to correct many semantic errors in model transformation programs.
This approach is based on evolutionary algorithms and test cases in the form of input/output models to find suitable patches to fix the transformation programs.
We discuss two limitations of EAs, namely single fitness peak and fitness plateaus, which are known to hinder the convergence of EAs approaches in this case and which make it difficult to find patches fixing three errors or more. 
To overcome these limitations, our approach is formulated as a multi-objective optimization problem and we use several objectives to guide the search. 
We dedicate an objective which gives a score based on the notion of social diversity that we defined on model differences.
We performed experiments to assess the impact of our approach, and especially on injecting social diversity in the process, on the effectiveness and the efficiency of repair approaches based on EAs and test cases.
Our results showed that injecting our social diversity measure in the search process improves both the effectiveness and the efficiency, and enables to find patches for faulty transformations having many errors.

\bibliographystyle{ACM-Reference-Format}
\bibliography{main.bib}

\end{document}